\documentclass{emulateapj}

\newcommand{\kms}{km s$^{-1}$}

\shorttitle{Arecibo Legacy Fast ALFA Survey IV.}
\shortauthors{Saintonge}

\begin{document}

\title{The Arecibo Legacy Fast ALFA Survey: IV. Strategies for Signal Identification and Survey Catalog Reliability}
\author{Am\'{e}lie Saintonge}
\affil{Center for Radiophysics and Space Research and National Astronomy and Ionosphere Center, Cornell University, Ithaca, NY, 14853, USA}
\email{amelie@astro.cornell.edu}

\begin{abstract}
We present a signal extraction utility written for the purposes of the Arecibo Legacy Fast ALFA survey (ALFALFA).  This survey, when completed, will have covered $\sim$7000~deg$^2$ of the high galactic latitude sky and should detect over 20,000 extragalactic objects.  It is the most sensitive blind HI survey to date. The large size of the survey justifies in itself the need for an automated way of identifying signals in the data set.  The matched-filtering signal extractor proposed is based on convolutions in the Fourier domain of templates of varying widths with each spectrum.  The chosen templates are built from a simple combination of Hermite functions to mimic the shape of typical galactic HI profiles of varying widths. The main advantages of this matched-filtering approach are a sensitivity to the total flux of the signals (and not only to peak flux), robustness against instabilities and short computing times. The details of the algorithm are given here, as well as results of simulations that assess the reliability and completeness of the process.
\end{abstract}
\keywords{methods: data analysis -- radio lines: galaxies -- galaxies: general -- catalogs -- surveys}

\section{Introduction}

In the past two decades, a number of all-sky astronomical surveys have been conducted throughout the electromagnetic spectrum, changing the way reseach is done. Endeavors such as the Sloan Digital Sky Survey in the optical \citep[SDSS;][]{stoughton}, the Two Micron All Sky Survey in the near-infrared \citep[2MASS;][]{2mass}, the Rosat All Sky Survey in X-ray \citep[RASS;][]{rosat}, to name a few, have provided a wealth of information about the extragalactic sky. In the radio wavelength regime however, even though 1.4 GHz (21cm) continuum surveys were conducted with synthesis telescopes (e.g. the NRAO VLA Sky Survey \citep[NVSS;][]{nvss}, the FIRST Survey \citep[]{first}), performing a large-scale 21cm line survey was an unconceivable project until the advent of multi-beam feed arrays. Some early HI 21cm surveys were carried out with the single pixel receivers at Arecibo during the 1990s, covering a few hundred square degrees each \citep[]{briggs,rosenberg}, indicating the need for such blind HI surveys but lacking the means of achieving significant sky coverage. Other blind HI surveys were made with several of the Green Bank telescopes \citep[]{fisher,krumm} and with synthesis telescopes \citep[e.g.][]{weinberg,deblok}. Even though these surveys achieved high sensitivity, they also seriously suffered from limited sky coverage.

Three radioastronomical facilities now have multi-beam L-band receivers that allow for large scale extragalactic surveys in the HI 21cm line. Using a thirteen beam array on the Parkes telescope, the HI Parkes All Sky Survey \citep[HIPASS;][]{hipass} was the first such survey and covered the declination range $-90^{\circ}<\delta<+25^{\circ}$ \citep[]{meyer,wong}. It was followed by the HI Jodrell All Sky Survey \citep[HIJASS;][]{hijass}, which uses a four element feed array and was intended to be the northern extension of HIPASS. Following a proposal by \citet[]{kildal}, a seven beam array named ALFA (Arecibo L-band Feed Array) finally became available at the Arecibo telescope in 2004. 

Making use of this instrument, the ongoing Arecibo Legacy Fast ALFA Survey \citep[ALFALFA;][]{g06a} is intended to cover 7000 deg$^{2}$ of high galactic latitude extragalactic sky. Observations began in early 2005, and should take five or six years to complete. Thanks to the large size of the Arecibo dish and the observing strategy implemented, ALFALFA provides data with eight times more sensitivity than HIPASS and four times better angular resolution. As highlighted in \citet[]{g06b}, the ALFALFA survey is much more sensitive to low-mass systems than the previous largest surveys, and will push the determination the faint end of the HI mass function to significantly lower masses than previous studies made possible \citep[]{zwaan97,rs02,zwaan03}. In combination with other surveys such as SDSS and 2MASS, ALFALFA will also address the issue of the ``missing satellites''. Other scientific objectives include the determination of the HI diameter function, the discovery of large HI tidal features and HI absorbers in the local Universe \citep[]{g06a}.

A challenge encountered in all the HI surveys mentioned above is to devise a technique to detect extragalactic signals out of the large data sets acquired. Independent galaxy-finding algorithms were created for HIPASS, HIJASS and ALFALFA. For the HIPASS data, the peak finding algorithm {\it MultiFind} was used \citep[]{kilborn}.   The algorithm searches for connected pixels above a user-defined peak flux threshold in individual velocity planes (which are cuts through a data cube at a given constant frequency).  The detections from successive planes are then compared, groupings of objects within 5 arcminutes of each other are made, and a candidate detection is identified when a source appears on at least two adjacent velocity planes.  The process is repeated for various degrees of smoothing along the spectral axis, and multiple detections are removed from the final catalog.  From simulated data cubes, {\it MultiFind} is found to detect $93\%$ of sources with peak flux greater than 60 mJy \citep[]{kilborn}.  When tested on real HIPASS data cubes however, {\it MultiFind} detects only $46\%$ of the galaxies visually detected, missing most of those with peak fluxes smaller than 50 mJy.  Therefore the algorithm could only be used to produce a catalog of the brightest galaxies in the sample \citep[]{koribalski}.

The other survey, HIJASS, used a slightly different approach where the signal detection process was done in two ways.  First, a list of candidates is made by visually inspecting the data cubes and recording objects found above the noise level that show both spatial and spectral extent \citep[]{hijass}.  Then, a second list of possible sources is made by applying an algorithm, {\it PolyFind}, which looks in the data cubes for peaks above a given threshold.  These signals are then compared to templates by matched-filtering and identified as potential sources if a good fit exists between signal and template.  The results of {\it PolyFind} are then checked by eye and the lists of detections from the visual and automated detections combined.

In the cases of both HIPASS and HIJASS, the automated signal extraction algorithms either rely strongly on visual inspection of the data or can only be used to detect the brightest objects.  For ALFALFA, a different strategy has been adopted in order to minimize the impact of these restrictions. Instead of identifying galaxies on the basis of their peak flux exceeding a defined threshold, a matched-filtering algorithm was conceived. Since performing the cross-correlations required by such a method are costly in computer time, the search is done in Fourier space using single-parameter templates made by a combination of Hermite polynomials and Gaussian functions. The advantages of this technique are many: the cross-correlation technique is sensitive to the total flux of a galaxy allowing for better sensitivity to broad and faint signals compared to peak-finding techniques, it acts robustly in presence of low frequency fluctuations of spectral baselines, and the execution time is reduced as the process does not need to be repeated for different degrees of spectral smoothing.

The remainder of the paper is structured as follows. In \S 2, we give a general description of the ALFALFA survey data, in \S \ref{1D} and \S \ref{3D} the workings of the signal extractor are explained and in \S 5 we present the results of simulations to quantify the performance of the extractor.

\section{ALFALFA Data \label{alfalfa}}
While a complete account of the ALFALFA data reduction pipeline ({\it LOVEDATA}) will be given elsewhere (Giovanelli et al. 2007, in preparation), we present here some of the key aspects that influence the process of signal extraction.

ALFALFA observations are done in drift mode, which means that on any given observing session the telescope is set at a specific declination and fourteen spectra (the seven beams of ALFA, each with two polarizations) are recorded every second as the sky drifts over the dish. A noise diode is fired every ten minutes to provide calibration. The ``drift map'' obtained over the course of the night is band-pass calibrated and baselined. Each drift scan is then examined by a member of the team to flag parts of the spectra that are contaminated by strong radio frequency interference (RFI). 

When the observations are completed over a region of the sky, all the drift scans affecting that region are combined to produce evenly gridded three-dimensional data cubes. These cubes are chosen to be $2.4^{\circ} \times 2.4^{\circ}$ in size, with grid points separated by 1\arcmin \ both in RA and Dec.. The third axis of the cubes, frequency, is made to have 1024 channels. In order to cover the full spectral range sampled by ALFALFA, four partly overlapping cubes are therefore made for each region on the sky, respectively covering the redshift ranges: -2000 to 3300\ \kms, 2500 to 7900\ \kms, 7200 to 12800\ \kms, and 12100 to 17900\ \kms. For each data cube, two different linear polarizations are separately recorded. Since the extragalactic signals are not polarized, the average of the two polarizations is used for most applications, but as explained in \S \ref{3D} it is useful to retain the information separately since a difference in flux between the two individual polarization cubes can be used as one of the criteria to reject spurious detections.

For each data cube, an array of identical size is created containing not the spectral data but a parameter describing the survey coverage at each position. When the data cubes are created, data sections flagged as affected by egregious RFI or of otherwise low quality are not used. However, low level RFI is not always easily identified and thus flagged, and can affect the gridded data and the signal extraction process.  Additionally, due to system malfunctions, parts of some drift scans may be missing, and therefore parts of some cubes sometimes have poorer coverage and lower sensitivity than others. This information is recorded in this additional cube and will also be used in the signal extraction process to reject unlikely detections.

\section{Fourier Space Matched-Filtering Signal Identification \label{1D}}

The signal extraction technique implemented for the ALFALFA survey relies on the cross-correlation of templates with the data. As outlined earlier, the motivation behind this approach is to obtain a catalog of galaxies that goes down to as low a flux limit as possible, while at the same time remaining efficient in terms of CPU and human time. In this section we first present the theory behind the matched-filtering technique used, and we then summarize its implementation. The process described in \S \ref{concept} is similar to that used by \citet[]{td79} to determine redshifts from galaxy spectra.

\subsection{Concept \label{concept}}
We want to find the best fit to a signal $g(x)$, where $x$ is the channel number that runs through all the frequencies sampled (as mentioned in \S 2, $x$ goes from 1 to 1024 for the adopted grid format).
We first assume that $g(x) \simeq \alpha t(x-\delta;\sigma)$, that
is the signal can be represented by a template function $t$, of which we can
change the amplitude $\alpha$, the central channel position $\delta$, and the width $\sigma$.
The best template will have values $\alpha$, $\delta$ and $\sigma$ such that 
\begin{equation}
\chi^2=\sum_{x=1}^{N}[\alpha t(x-\delta;\sigma)-g(x)]^2, \label{chi2}
\end{equation}
is minimized, where $N=1024$ is the last spectral channel.
Now, let $\sigma_g^2$ and $\sigma_t^2$ be the variance of the signal and template, respectively
\begin{eqnarray}
\sigma_g^2 & \equiv & \frac{1}{N} \sum_{x}g(x)^2 \label{sigmag} \\
\sigma_t^2 & \equiv & \frac{1}{N} \sum_{x}t(x)^2, \label{sigmat}
\end{eqnarray}
and let $c(x)$ be the normalized cross-correlation function,
\begin{equation}
c(x)=g(x) \ast t(x-\delta;\sigma) = \frac{1}{N\sigma_g\sigma_t}\sum_{n}[g(n)t(n-x)]. \label{cross}
\end{equation}
Expanding equation \ref{chi2} and including these quantities, $\chi^2$ can be written as:
\begin{equation}
\chi^2 = \alpha^2 N \sigma_t^2 + N \sigma_g^2 - 2 \alpha N\sigma_g \sigma_t c(\delta). \label{chisimple}
\end{equation}

This expression for $\chi^2$ has to be minimized with respect to $\alpha$, $\delta$ and $\sigma$ in order to find the best-match template for the signal $g(x)$.  
First, equation \ref{chisimple} can be minimized with respect to $\alpha$:
\begin{equation}
\frac{\partial \chi^2}{\partial \alpha} = 2 \alpha N \sigma_t^2 - 2N \sigma_g \sigma_t
c(\delta).
\end{equation}
Equating this to 0 gives a simple expression for the template peak height as a function of the cross-correlation function and the standard deviations of the spectrum and template functions:
\begin{equation}
\alpha = \frac{\sigma_g}{\sigma_t}c(\delta). \label{amin}
\end{equation}
This expression for $\alpha$ can now be introduced in equation \ref{chisimple} to give
\begin{equation}
\chi^2 = N \sigma_g^2 [1-c(\delta)^2], \label{chifinal}
\end{equation}
which shows that minimizing $\chi^2$ is equivalent to maximizing $c(\delta)$, the value of the cross-correlation between the data and the 
template of width $\sigma$ centered on the spectral channel $\delta$.

\subsection{Implementation}

Because time efficiency is an issue, the signal search is not done in the radio frequency domain since calculating the summations of equation \ref{cross} can be very time-consuming.
It is well known that convolutions are most efficiently implemented by using the Fast Fourier Transform algorithm and invoking the convolution theorem. Let $G(k)$ and $T(k)$ be the discrete Fourier transforms of the spectrum and template,
\begin{eqnarray}
G(k) &=& \sum_{x} g(x) e^{-2 \pi ixk/N} \label{FTg}\\
T(k) &=& \sum_{x} t(x) e^{-2 \pi ixk/N}. \label{FTt}
\end{eqnarray}
Then, by virtue of the convolution theorem, the Fourier transform of the cross-correlation function is
\begin{equation}
C(k) = \frac{1}{N \sigma_g \sigma_t} G(k) T^{\star}(k), \label{FTc}
\end{equation}
where $^{\star}$ indicates the complex conjugation operation. The function $c(x)$ is simply retrieved by taking the inverse discrete Fourier transform of $C(k)$.

Putting all the pieces together, the process of signal extraction goes as follows. A set of templates covering a physical range of rotation widths is selected (the specific shape of these templates is described in \S \ref{templates}), and their Fourier transforms calculated. Then, the Fourier transform of each spectrum is multiplied with each of the templates following equation \ref{FTc}, and the inverse transform of this product computed. We thus have a set of convolution functions, one for each template. For each of these convolution functions, the maximum value $c(\delta)$ is determined, where $\delta$ is the spectral channel at which $c(x)$ is maximized. We are then left with a set of values of $c(\delta)$, one for each template. The largest of these values, $c(\delta_{max})$, tells us which template had the most appropriate width to model the signal. Two of the parameters of the model are thus already determined: the central position, $\delta_{max}$, and $\sigma_{max}$, the width of the template that produced $c(\delta_{max})$). The third parameter, the peak amplitude $\alpha_{max}$, is simply calculated using equation \ref{amin}.

This last statement outlines one of the strengths of this method: by performing a very small number of cross-correlations with carefully chosen templates, the parameters describing a best-fit model $\alpha t(x - \delta;\sigma)$ to a signal $g(x)$ can be accurately determined with a single free parameter, the spectral width.

\subsection{Templates \label{templates}}

We have so far not specified what is the shape of the adopted templates. For the above strategy to apply, the selected templates must be characterized by a width, maximum amplitude and central position along the spectrum. The first two obvious choices are simple Gaussian and top-hat functions. The Gaussian has the advantage of being simple to manipulate and normalize and is very adequate to model galaxies of small rotation widths. However, for broader signals exhibiting the characteristic two-horn profile of extragalactic HI spectra, a Gaussian provides a poorer fit and therefore is likely to miss very broad and shallow signals. A top-hat template would be less affected by this inadequacy, as its shape better approximates that of broad HI signals. A possible approach would be to use a hybrid template, one for example in which the narrow templates are Gaussian and the broader ones top-hats, or templates made of two Gaussians of increasing width and spectral separation. There are however problems with these two possibilities. The Gaussian/top-hat templates would suffer from the discontinuity between the two models, and both would prove difficult to normalize while keeping track of the template width. This last point is important since the values of the cross-correlation between the spectrum and all the different templates need to be directly compared.

The solution to this problem is to find a set of templates with shapes that evolve with width, but that are at the same time a homogeneous family of functions so the value of the cross-correlation can be compared from one template to the next in order to find the best match for any given signal. Ideally, this family of functions would contain a Gaussian as its lowest order term since it is a very appropriate template for narrow signals.  At the same time, it should be able to describe a top-hat-like function for the broader signals. 

We therefore introduce a template model with a shape which evolves smoothly with width. To make the width-dependence tractable, we express the templates in the functional basis formed by the Hermite functions, $\Psi_n(x)$. These functions are the product of the Hermite polynomials, $H_n(x)$, with a Gaussian function and are most commonly known as the eigenfunctions of the one-dimensional quantum harmonic oscillator. This family of functions is a convenient choice since $\Psi_0(x)$ is a Gaussian and because, owing to their orthogonality, any template can easily be expressed as $\sum_{i} a_i \Psi_i(x)$, where $a_i$ carries the dependency on the width $\sigma$.

\begin{figure}[ht!]
\epsscale{0.8}
\plotone{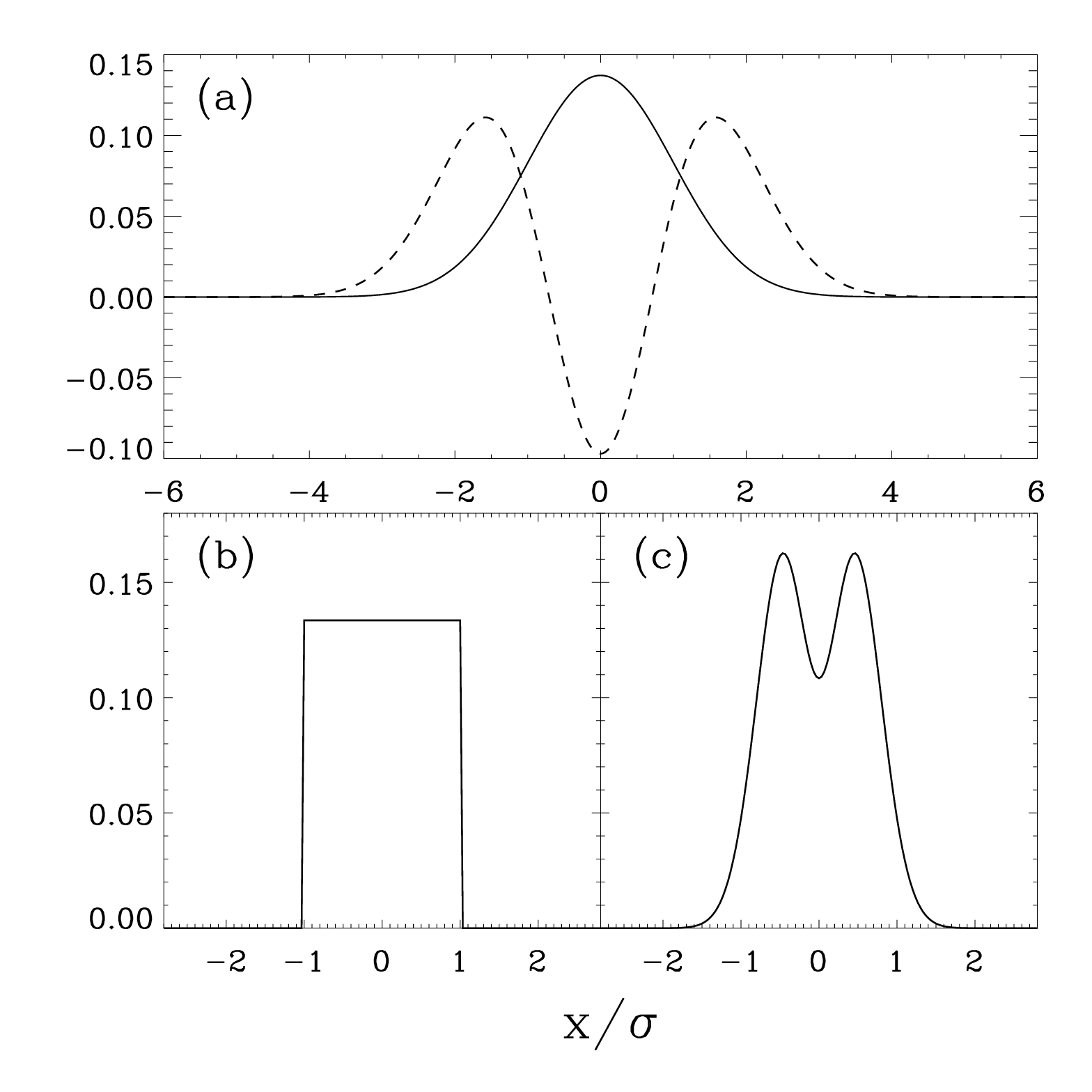}
\caption{(a) The first two symmetric Hermite functions, $\Psi_0(x)$ (a Gaussian function, the continuous line) and $\Psi_2(x)$ (dashed line).  (b) Example of a possible template, a top-hat function.  (c) The expansion of the top-hat function in terms of $\Psi_0(x)$ and $\Psi_2(x)$.  This template very conveniently shows the two-horn shape of most HI profiles. \label{herm_fncs}}
\end{figure}

Since the templates should be symmetric around the central velocity, let's consider the first two symmetric Hermite functions:
\begin{equation}
\Psi_0(x;\sigma)=\frac{1}{\sqrt{\sigma \pi^{1/2}}}e^{-(x/\sigma)^2/2}
\end{equation}
\begin{equation}
\Psi_2(x;\sigma)=\frac{1}{\sqrt{\sigma \pi^{1/2}}} \left(\frac{-1}{\sqrt{2}}+\sqrt{2}\left(\frac{x}{\sigma}\right)^2\right) e^{-(x/\sigma)^2/2}
\end{equation}
These two functions are plotted in panel (a) of Figure \ref{herm_fncs}. 
Since the first term is a pure Gaussian, the narrowest templates can be taken to be the first Hermite function alone. For broader signals, the top-hat function can be expanded in terms of the above two functions by performing simple scalar products. We first create a top-hat function, $c(x)$, and expand it in terms of $\Psi_0(x;\sigma)$ and $\Psi_2(x;\sigma)$:
\begin{equation}
c_H(x)=\frac{a_0\Psi_0(x)+a_2\Psi_2(x)}{\sqrt{a_0^2+a_2^2}},
\end{equation}
where $a_0=\Psi_0(x)\cdot c(x)$ and $a_2=\Psi_2(x)\cdot c(x)$.
See panels (b) and (c) of Figure \ref{herm_fncs} for an illustration. Note that $c(x)$ could be expanded using an infinity of terms, but keeping the first two is sufficient to get the desired template shape.

Then we define two different width thresholds, $\sigma_{T1}$ and $\sigma_{T2}$. The templates having $\sigma<\sigma_{T1}$ will be pure Gaussians, while the ones with $\sigma_{T1}<\sigma<\sigma_{T2}$ will be a hybrid between the pure Gaussian and the expansion of the top-hat.  The templates wider than $\sigma_{T2}$ will be $c_H(x)$ stretched or compressed to the appropriate width. Explicitly, the templates have the functional form: 
\begin{equation}
t(x;\sigma)=\Bigg\{\footnotesize \begin{array}{ll}
\Psi_0(x;\sigma) & \mbox{if $\sigma<\sigma_{T1}$}\\
      b_f\Psi_0(x;\sigma) + \sqrt{1-b_f^2}\Psi_2(x;\sigma) & \mbox{if $\sigma_{T1}<\sigma<\sigma_{T2}$}\\
      a_{N,0}\Psi_0(x;\sigma) + a_{N,2}\Psi_2(x;\sigma) & \mbox{if $\sigma>\sigma_{T2}$}
     \end{array}
\label{templ_func}
\end{equation}
where $b_f=f\left(1-a_{N,0}\right)+a_{N,0}$, $0<f<1$, and $a_{N,i}=a_i\sqrt{a_0^2+a_2^2}, (i=0,2)$. From experience of measuring HI profiles of galaxies of various rotation velocities, the values of the two thresholds are set such that FWHM$\left(t(x;\sigma_{T1})\right)=60$ \ km s$^{-1}$ and FWHM$\left(t(x;\sigma_{T2})\right)=200$ \ km s$^{-1}$.

\begin{figure}[ht!]
\epsscale{0.8}
\plotone{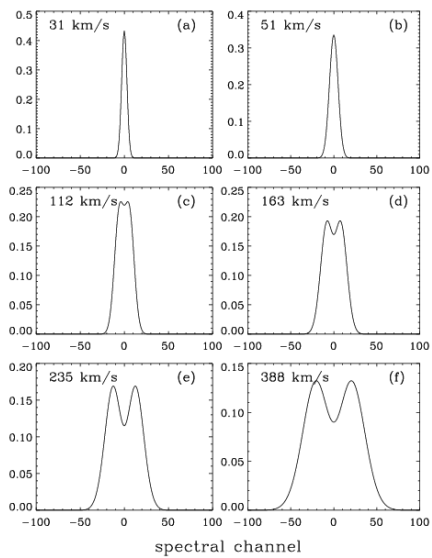}
\caption{Example of templates created by the functions presented in equation \ref{templ_func}.  The velocity in the top left corner of each panel is the FWHM of the template.  The top two panels are pure Gaussians, in the second row are represented two templates of the ``hybrid'' type and in the bottom row are two templates made from the expansion of the top-hat function. \label{templ_plot}}
\end{figure}

The advantage of using a set of orthogonal functions is that the normalization of the various templates is trivial, making for a smooth transition from one width interval to the next, a criterion that is essential for the amplitude of the cross-correlation functions to be compared. In Figure \ref{templ_plot} we present six examples of templates. The velocity in the top left corner of each panel is the FWHM of each template. The top row represents the $\sigma<\sigma_{T1}$ case, the middle line the $\sigma_{T1}<\sigma<\sigma_{T2}$ and the bottom row the widest templates which have $\sigma>\sigma_{T2}$. These templates provide representation of real signal which are not perfect, but approximate real data quite adequately.

\subsection{Validation \label{validation}}

\begin{figure}[ht!]
\epsscale{0.8}
\plotone{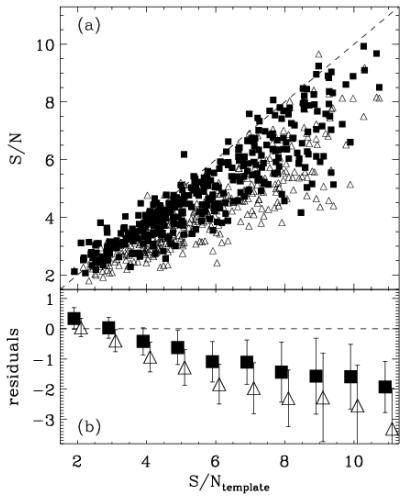}
\caption{Comparisons of different techniques that can be used to perform the signal extraction. (a) Comparison of the $S/N$ ratio measured using the ``Hermite templates`` matched-filtering technique (filled squares) and a boxcar smoothing and peak-finding method (open triangles) with the optimal result obtained by convolving the simulated profile with itself (identified as $S/N_{template})$. (b) Residuals of the data shown in (a) from a 1:1 relation.
\label{ex3d_box}}
\end{figure}

In order to test this choice of template and method, simple simulations are performed with three different signal extraction techniques: (1) the matched-filtering process described above with the Hermite function templates of Figure \ref{templ_plot}, (2) a boxcar smoothing of the data followed by a peak search, and (3) the matched-filtering process but this time with the simulated galaxy profile serving as the template. This last technique is used as a reference, since searching for a signal with itself produces optimal results. The results are shown in Figure \ref{ex3d_box}, and represent 500 simulated spectra in each of which a spectral profile was injected using the templates  described in \S \ref{sim_galaxies}. The top pannel compares the results of techniques (1) and (2) (filled squares and open triangles, respectively) with that of technique (3) (labeled as $S/N_{template}$ in the figure). As expected, none of the two techniques can recover all of the signal like the template itself can, especially for the fast-rotating galaxies which tend to have more complicated spectral profiles. However, there is a significant difference between the two, with the ``Hermite templates'' matched-filtering technique producing superior results than the boxcar smoothing method. Moreover, while a boxcar smooth and peak-finding search seems like a conceptually simpler technique, it turns out to be more complicated to implement and take more computation time due to the need to repeat the process for several degrees of smoothing. Therefore we conclude that while the two techniques are mathematically similar, the method proposed here is superior to the boxcar approach because of the specific choice of the template shapes, and the Fourier space technique used to implement it. It is also more resilient against broad-band baseline fluctuations.

\section{Three-dimensional Extraction : Adding Spatial Information \label{3D}}

The process described in the previous section is used to detect signals one spectrum at the time. However since the grid points in an ALFALFA data cube are separated by 1\arcmin \ while the size of the Arecibo beam is $\sim3.5$\arcmin, a real detection is expected to be made in several 
spectra. Ideally, the signal extraction process would be performed simultaneously in the three directions (frequency, RA and Dec) by convolving 3-dimensional templates with the Fourier transform of the whole data cube. This method would be most sensitive, but is unfortunately not practiced due to the large size of the arrays. This is why the search is done one spectrum at a time in Fourier space, with the next step of the extraction aimed at matching detections from different spectra that correspond to the same HI source. 

In the first phase of the process (the 1-D signal extraction along the frequency axis of the data cube, see \S \ref{1D}), an array of identical size to the data cube itself is created with value zero everywhere except at the coordinates and central spectral channel of each of the detections. The next phase is iterative. The maximum value within this array is found: this corresponds to the $S/N$ value at the center of the brightest HI source within the data cube. A small box centered on this position is then defined to contain all spectral detections that may be part of this galaxy. All the points in this box are projected in a single RA-Dec plane and a two-dimensional Gaussian is fitted to determine the spatial extent of the galaxy. Thus, a candidate detection is obtained and logged. Afterwards, all grid points in the array associated with that source are set to zero, and the second highest $S/N$ source is sought. The process repeats itself until the highest S/N value left in the grid is below the user-defined $S/N$ detection threshold. Note that we reject any detection made in the first or last 10 channels of the cube, to avoid the possible spurious detections that could come from the fact that the FFT algorithm assumes that the signal is periodic.

At this point, the spatial information on each source and that contained in the two individual polarization cubes can be used to reject spurious detections. Since extragalactic HI signals are not polarized, the flux of a galaxy in each polarization should be statistically the same, while the noise should be uncorrelated between the two. Therefore a criterion that will reject a detection if there is a significant flux difference between the two polarizations can be applied. The power of this test is limited at low $S/N$ by the fact that the presence of noise with near-Gaussian properties will create a difference in flux even for a perfectly unpolarized signal. The polarization criterion is useful for bright signals and to reject bad detections arising from strong RFI, but it can only be taken as an indicator of caution for the fainter detections, where it would be most useful.

A number of other criteria are applied to select candidate detections. They are based on the knowledge of the expected extent of HI disks in galaxies, of the general properties of spectra in the 21cm line, and of the effect of the Arecibo beam on the observed properties of galaxies.  A detection will be rejected if:
\begin{itemize}
\item the FWHM of the detection is larger than 15\arcmin \ along the major axis (the relatively small number of objects of large size in the sky are easily spotted by visual inspection),
\item no spatial extent in the ALFALFA grid is measured in either RA or Dec, 
\item the detection is made in fewer than four grid points,
\item the single-spectrum contiguous detections assigned to the same source have significantly different velocity widths,
\item the survey sky coverage at this spatial position is incomplete, indicating poor data quality.
\end{itemize}

The sources passing those tests are considered candidate detections. At this point, the user can quickly browse through the data cube and either add to or remove from this list. A catalog of candidate detections is finally produced, containing the following parameters for each source: HI position, redshift, velocity width, HI diameters, peak flux, flux integral, noise rms and $S/N$ ratio. Specialized measuring algorithms which allow, for example, baseline fitting, deblending and more detailed characterization of signal parameters, can be applied interactively to the candidate detections delivered by the signal extractor.

\section{Simulating ALFALFA Data \label{simulations}}

To test the performance of the signal extraction process, simulations were made. The main goal of this exercise was to establish the completeness and reliability as a function of $S/N$ for the catalogs of candidate detections, by simulating datasets with properties mimicking those of the ALFALFA data cubes. First, arrays of Gaussian noise are produced and smoothed to the resolution of the survey data. Then galaxy spectral profiles of varying spatial size and spectral characteristics are injected into these grids of noise arrays. Finally we carry on the signal extraction process and compare the output catalog with the input information of the simulated galaxies. Below we describe those steps in detail.

\subsection{Noise Characterization \label{noise}}

\begin{figure}[ht!]
\plotone{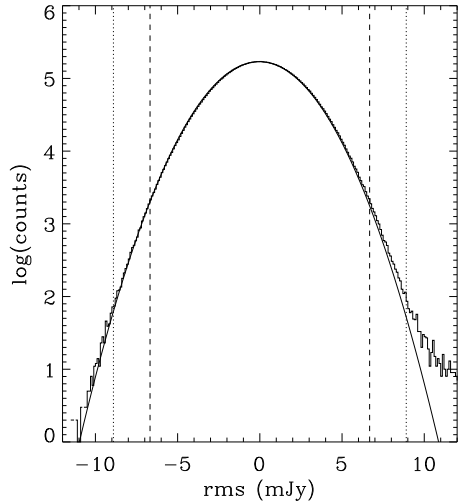}
\caption{Histogram of pixel values for ALFALFA data cubes after smoothing to a spectral resolution of 10 \kms.  The solid line represents the best Gaussian fit to the histogram.  The standard deviation of the noise distribution is $\sigma=2.226\pm0.001$ mJy. The vertical dashed lines are at $\pm3\sigma$ and the dotted lines at $\pm4\sigma$.\label{noise_hist}}
\end{figure}

The characteristics of noise in the ALFALFA data cubes greatly impact the performance of the signal extractor. Therefore, before attempting to simulate data, we first analyze the noise present in the ALFALFA survey datacubes. In Figure \ref{noise_hist} we present the histogram of pixel values for a set of ALFALFA data cubes, after smoothing to a spectral resolution of 10 \kms, excluding spectral channels containing Galactic HI. The continuous line is the best Gaussian fit to the histogram, which has a standard deviation of $\sigma=2.226\pm0.001$ mJy. The vertical dashed lines are at $\pm3\sigma$ and the dotted lines at $\pm4\sigma$. The rms of the residuals of the fit is 0.01 dex within $\pm3\sigma$ and 0.04 dex within 
$\pm4\sigma$. This means that the pixels within $\pm3\sigma$ deviate from Gaussian white noise by less that $1\%$. The data cubes used to create the histogram contained galaxies, the contribution of which can be seen above the $3\sigma$ level where they start to dominate over the noise. The idea that the deviation from Gaussianity for positive pixel values comes from galactic contributions is supported by the fact that the negative pixels follow the Gaussian distribution down to the $5\sigma$ level. The simulations are therefore performed assuming Gaussian white noise with $\sigma=2.2$ mJy.

Data cubes of size $2.4^{\circ} \times 2.4^{\circ}$ with grid points separated by 1\arcmin \ and 1024 spectral channels are created out of Gaussian white noise with the above characteristics. The cubes are then smoothed by convolving a two-dimensional Gaussian with a 2\arcmin \ HPFW kernel and by performing 3 point Hanning smoothing along the spectral direction, to reproduce the 10 \kms \ resolution of the ALFALFA data cubes. These cubes are the basic units in which simulated galaxies are then injected.

\subsection{Simulating Galaxies \label{sim_galaxies}}

To simulate galaxies, we use as models the 166 galaxies detected during the ALFALFA precursor observations \citep[]{g06b}.  This way, the extractor is tested on actual profiles and its performance can be assessed most reliably.  In each data cube 40 simulated galaxies are introduced while making sure that each galaxy is separated from its nearest neighbour by at least 20\arcmin \ and 200 km/s$^{-1}$. For each galaxy, the following parameters are randomly assigned: central spectral channel, right ascension and declination pixel coordinates, peak flux, velocity width and physical sizes along the major and minor axes.  One of the 166 models is then randomly chosen, scaled to the assigned peak flux, stretched or compressed to the selected spectral width, and added to the grid at the randomly selected position.  Since sources appear in multiple grid points, the model galaxy is also added to all nearby grid points after scaling it down assuming that the integrated flux drops off from the central position according to a two-dimensional Gaussian profile.  We simulate the second polarization in an identical way, except that the template is scaled up or down by up to $5\%$ to simulate possible offsets in calibration, thought to be representative of the ALFALFA data calibration errors. For completeness we repeat the process by simulating galaxy spectra as Gaussians or as top-hat models. The performance of the signal extractor is not affected by this change of simulated spectral profile. 

\subsection{Signal Extractor Performance}

\begin{figure}[ht!]
\plotone{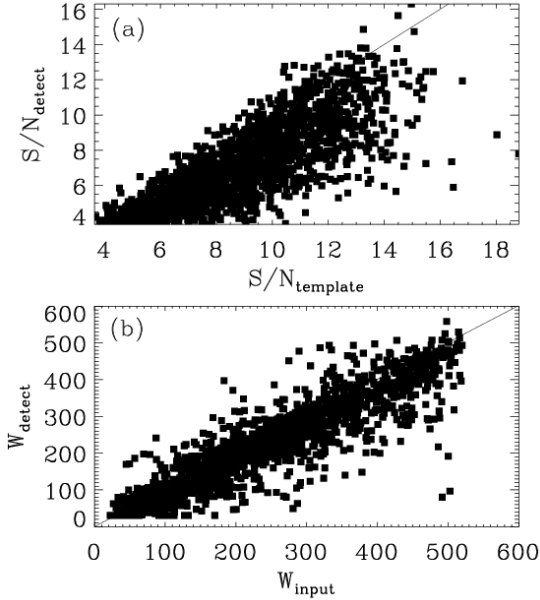}
\caption{Comparison of the parameters calculated by the signal extractor with the input of the simulation.  Represented are 1800 simulated galaxies detected by the signal extractor.  (a) S/N ratio values of the simulated galaxies ($S/N_{template}$) and those measured by the signal extractor ($S/N_{detect}$). (b) rotation width of the simulated galaxies ($W_{input}$) and those measured by the signal extractor ($W_{detect}$).  In both pannels the solid line is the 1:1 relation. \label{params_ex}}
\end{figure}

As a first test, the input parameters of the simulated galaxies are compared to those recovered by the signal extractor. In Figure \ref{params_ex}, the input and ``detected'' values for the velocity width $W$ and $S/N$ ratio are plotted against each other, the line being the 1:1 relation.  The overall agreement between input and output parameters is quite satisfactory, considering that the goal of the signal extractor is to detect galaxies, but not necessarily to produce the most accurate values of their parameters, something that is achieved during the next step of the data processing pipeline (Giovanelli et al. 2007, in preparation).  As in section \ref{validation}, $S/N_{template}$ is the maximum $S/N$ that could be retrieved by the signal extractor, as it was calculated using the simulated galaxy profile itself as the matched-filter. Note that here and in all that follows, the $S/N$ ratio is defined as the mean flux across the signal to rms ratio, if the spectrum was smoothed to half the velocity width of the signal, up to a width of 400 \kms (where typical baseline fluctuations start to be of the same width as the galaxy profiles):
\begin{equation}
S/N=\Bigg\{ \begin{array}{ll}
\frac{F/W}{\sigma} \times \left( \frac{W/2}{10 {\rm \ km \ s}^{-1}} \right)^{1/2} & \mbox{if $W<400$ \kms}\\
    \frac{F/W}{\sigma} \times \left( \frac{400/2 {\rm \ km \ s}^{-1}}{10 {\rm \ km \ s}^{-1}} \right)^{1/2} & \mbox{if $W\geqslant400$ \kms,}
     \end{array}
\label{sn}
\end{equation}
where $F$ is the integrated flux in mJy \kms, $W$ is the width of the signal in \kms \ and $\sigma$ is the rms noise.
In the above equation, 10 \kms \ corresponds to the spectral resolution of the ALFALFA data, after Hanning smoothing. This definition of $S/N$ takes into account the fact that for the same peak flux a broader signal has more signal, something the signal extractor is sensitive to.

\begin{figure}[ht!]
\plotone{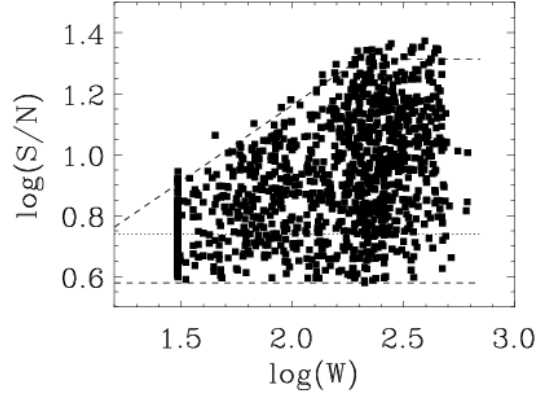}
\caption{Distribution of the 1150 simulated galaxies detected by the signal extractor in the $\log(W) - \log(S/N)$ plane.  The bottom dashed line is the $S/N$ threshold of 3.8 used during signal extraction, and the top dashed line the maximum $S/N$ ratio simulated for any given width.  The dotted line is $S/N=5.5$, which is where the catalog of detections becomes completely reliable. The feature seen at $W=30$ \kms \ occurs because this specific value of $W$ corresponds to the narrowest template used in the signal extraction process. \label{sn_w}}
\end{figure}

In Figure \ref{sn_w}, we plot the $S/N$ ratio of the detected galaxies against their velocity width to check if the catalog is biased against either narrow or broad signals: the top dashed line is the maximum $S/N$ ratio of the simulated data as injected, the bottom dashed line is the $S/N$ threshold of 3.8 used during the signal extraction process and the dotted line corresponds to the $S/N$ level of 5.5, where the catalog of detections becomes reliable (see \S \ref{reliability}). The region of the plot between the bottom dashed line and the dotted line is under-populated, especially in the region $\log(W)>2.5$, but all candidate sources ``detected'' above the dotted line actually correspond to sources injected in the simulation. The crowding in the plot at $W=30$ \kms \ reflects the width of the narrowest template used in the signal extraction process.

\subsection{Completeness and Reliability \label{reliability}}

\begin{figure}[ht!]
\plotone{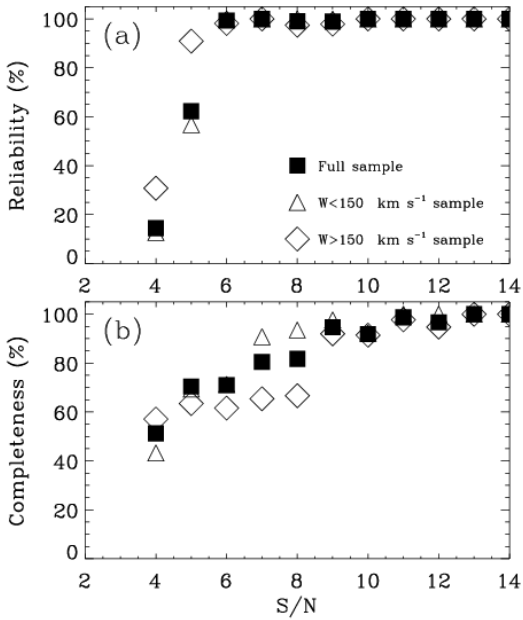}
\caption{Reliability ({\it the fraction of detections that are simulated sources}) and completeness ({\it the fraction of the simulated sources that are recovered by the extractor}) of the catalog produced by the signal extractor, based on the simulation of 1500 galaxies.  The filled squares represent the whole sample, the open triangles the galaxies with $W<150$\kms and the open diamonds those with $W>150$\kms. (a) Reliability: as a function of $S/N$, fraction of the detections made that correspond to simulated galaxies. (b) Completeness: fraction of the simulated galaxies that are detected, as a function of $S/N$. \label{compl_rel}}
\end{figure}

The most important goal of performing simulations is to assess the reliability and completeness of the catalogs produced by the signal extractor. We define ``reliability'' as the fraction of candidate sources detected by the extractor which correspond to sources actually injected in the simulations. The complement to 1 of the reliability fraction refers to spurious detections. By ``completeness'', we mean the fraction of the simulated sources that are recovered by the extractor.

These issues are critical because they influence the survey strategy. The ALFALFA catalog is built in two stages. A first catalog is made with the detections having $S/N>S/N_{conf}$, where $S/N_{conf}$ is the threshold above which the probability of a detection being real is extremely high. The second step will require re-observation of the detections with $S/N_{prob}<S/N<S/N_{conf}$, where $S/N_{prob}$ is the limit below which most detections are spurious and the process of re-observation would yield too few confirmations to justify the extensive telescope time required to carry the observations. 

By studying the reliability of the simulated catalog, the values of $S/N_{prob}$ and $S/N_{conf}$ can be determined. The top panel of Figure \ref{compl_rel} shows as a function of $S/N$ the fraction of detected sources matched by input signals in the simulations; in each $S/N$ bin the number of simulated sources detected is divided by the total number of detections. The filled squares represent the full sample of detected galaxies, the open triangles those with $W<150$ \kms and the open diamonds the detections with $W>150$ \kms. The reliability figure shows that any detection with $S/N>5.5$ is reliable. For detections in the $4.5<S/N<5.5$ range, $91\%$ of them are real for $W>150$ \kms but only $57\%$ for $W<150$ \kms. Detections with $3.5<S/N<4.5$ are much less reliable with an overall probability of $14\%$ that any detection is real. In the bins centered on $S/N=4$ and $S/N=5$, detections of broader signal are more reliable than those of narrower signals. This is a reasonable result since the Gaussian noise is uncorrelated between spectral channels past the 3 point Hanning smoothing level, and therefore spurious detections are much more likely to be very narrow features than broad ones. Considering these results, the adopted values for the two thresholds are $S/N_{conf}=5.5$ and $S/N_{prob}=4.0$

The bottom panel of Figure \ref{compl_rel} shows the completeness of the extracted catalog as a function of $S/N$ for the sample of simulated galaxies. 
We plot the fraction of sources injected in the simulations actually detected by the extractor. The symbols refer to the same sample of galaxies as in the top panel. The catalog of sources produced by the signal extractor is complete to better than $50\%$ for $S/N>4.0$ (however, the vast majority of candidate detections at $S/N=4.0$ are spurious). For the narrowest galaxies, the catalog becomes complete to better than $90\%$ at $S/N>6.5$, while this level of completeness is reached at $S/N>8.5$ for the widest sources.

\begin{figure}[ht!]
\epsscale{0.8}
\plotone{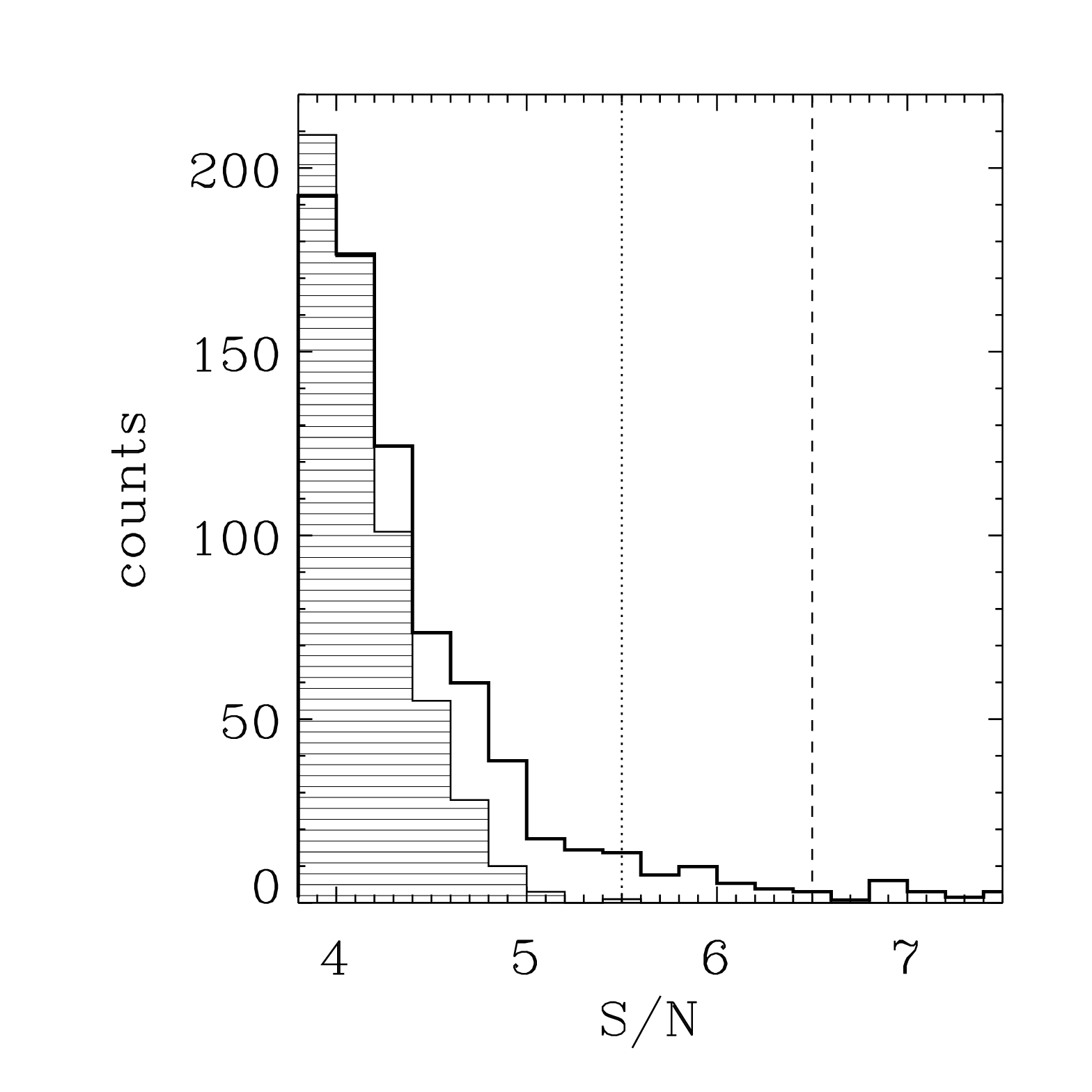}
\caption{$S/N$ distribution of the detections made by the signal extractor on purely noisy data. The filled histogram is the results for simulated cubes of Gaussian noise (see \S \ref{noise}), while the open histogram is for the ``galaxy-free'' spectral range of real ALFALFA data cubes. The dotted line corresponds to $S/N_{conf}=5.5$ as determined from the reliability curve of Figure \ref{compl_rel}, but the dashed line represents the adopted value of $S/N_{conf}=6.5$ after taking into account the additional contributions to the noise that were not simulated.
\label{spurious}}
\end{figure}

All the above results were obtained by simulating ``ideal'' data cubes. While uniform Gaussian white noise is present over most parts of the spectra, some frequency ranges and some parts of the sky coverage are affected either by residual RFI or by the edges of the bandpass where the noise rms is locally higher or the noise features start to be correlated.  The presence of standing waves produced by strong continuum sources and system instabilities also affect the quality of the baselines. The signal extractor is fairly robust against these problems since it will reject detections if the two polarizations do not match, calculates the noise rms over nearby pixels only, and is insensitive to large-scale amplitude variations of the baseline larger than the maximum template width. The thresholds defined above will vary depending on data quality. 

We perform an additional test to verify the influence of data quality on the detection process. Since the ALFALFA velocity coverage starts at -2000 \kms \ while no galactic or extragalactic HI signals are expected to be found in the -2000 to -500 \kms \ velocity range, that spectral region has the characteristics of the survey data while being ``uncontaminated'' by the presence of cosmic sources, except for the very rare possibility of OH Megamasers. We compare the detections made by the signal extractor in that velocity range for 120 ALFALFA data cubes to the detections coming from cubes of simulated noise (as explained in \S \ref{noise}).

Histograms showing the $S/N$ distribution of these detections are presented in Figure \ref{spurious}. The filled histogram represents the simulated data sets and the open histogram the ``galaxy-free'' portion of the ALFALFA data cubes. The two histograms are normalized to represent the same number of independent spectra processed. For each detection made in the ALFALFA data cubes, a weight is defined to characterize the data quality. The values range from 0 (no data) to 1 (perfect survey coverage). The open histogram shows all the detections located in regions that have a weight larger than 0.55. This value is adopted to provide the best match between the two histograms for $S/N<4.5$ where the Gaussian noise component dominates. Above $S/N=4.5$, there is an excess of candidate detections from the ALFALFA data cubes over that of simulated data characterized by Gaussian noise. This disagreement can be completely eliminated by allowing in the sample only sources found in regions with weights larger than 0.9, but this criterion would also reject a sizeable fraction of the survey volume. The excess is believed to be caused by residual RFI, probably of low intensity level, which the flagging stages of the data processing failed to identify. Much of the residual RFI is found in regions of the volume sampled by the survey which are in or in the vicinity of regions of low weight. Low level RFI may however be present, and largely unidentified, in regions of high weights. We thus relax the definition of $S/N_{conf}$ from the value of 5.5, as derived from the discussion related to Figure \ref{compl_rel}, to the value of 6.5 (dashed line in Figure \ref{spurious}). We estimate that at $S/N>6.5$, spurious features will appear in the data -- and be tentatively detected by the signal extractor -- at the rate of one every six square degrees. This translates into a reliability of $\sim95\%$ that candidate detections with $S/N>6.5$ correspond to real cosmic sources. Visual inspection of the sources may reveal characteristics that help further increase the reliability figure.

As for $S/N_{prob}$, we would like to keep a catalog of low $S/N$ candidate sources having a likelihood of being associated with cosmic sources such that follow-up corroborating observations can detect them at a rate, per unit of telescope time, which is significantly higher than that at which ALFALFA blindly detects sources with $S/N>S/N_{conf}$. Estimating that the latter number is of order 6 sources per hour (see \citet[]{g06a}), and that a corroborating observation may require 2.5 minutes of telescope time (24 pointed observations per hour), it appears as if $S/N_{prob}$ can be set at a level such that the estimated reliability is of order $25\%$. Inspection of Figures \ref{compl_rel}a and \ref{spurious} would indicate that $S/N_{prob}=4.5\pm0.5$: sources with $S/N>S/N_{conf}=6.5$ can be considered fairly safe detections, better than $95\%$ reliable; sources with $S/N_{prob}=4.5<S/N<S/N_{conf}=6.5$ would be kept in a separate catalog to be used for pointed, corroborating observations after completion of the survey.

\section{Summary}
We have described a new signal extraction algorithm created for the ALFALFA survey, which is a major on-going effort at the Arecibo Observatory to blindly map in the HI 21 cm line 7000 deg$^2$ of extragalactic sky. All the software needed for the data reduction pipeline, known as the {\it LOVEDATA} package, was created in the IDL environment by members of the ALFALFA collaboration. Part of this package is a utility that automatically finds HI sources in the ALFALFA data cubes and creates catalogs of candidate detections. The need for such a tool first stems from the sheer size of the project, and is justified by the desire to create a uniform catalog of sources, even though the data processing and analysis is carried on by a large number of individuals from multiple institutions.

The strategy adopted for the ALFALFA signal extractor is that of cross-correlations of templates with the spectra. This matched-filtering technique, when performed in Fourier space, has the advantage of being very time-efficient. Other advantages include the fact that the process is sensitive to the total flux of the galaxies, rather than to the peak flux, as is the case for algorithms that work using a peak-finding strategy, and the fact that the algorithm is quite robust against baseline and other instabilities. The templates used are built with the first two symmetrical Hermite functions, which are the product of the Hermite polynomials with a Gaussian function. The narrowest templates with widths smaller that 60 \kms \ are simple Gaussians, while the wider templates are built via a truncated expansion of a top-hat function in terms of the Hermite functions. This way the templates used for narrow signals have a Gaussian shape, while the wider ones show the characteristic two-horned profile of most HI emission from rotating galaxies. This allows for an optimal detection rate of galaxies of all rotation widths, while having a single-parameter (the velocity width) set of templates that are all constructed from the same family of functions. As explained in \S \ref{1D}, this condition is required for the Fourier space cross-correlation technique to apply efficiently.

A set of simulations were performed to assess the performance of the signal extractor. Following an analysis of the noise in the ALFALFA data cubes, simulated cubes were created and modeled galaxies pasted into them. The signal extractor recovers the velocity width and $S/N$ ratio of the input galaxies, and is not biased against either narrow or broad signals. The simulations suggest that, in the presence of Gaussian noise, detections with $S/N>5.5$ are reliable to better than $98\%$ confidence, and broad detections are reliable up to the $91\%$ level at $4.5<S/N<5.5$. The presence of low level RFI inadequately flagged during the data processing reduces the reliability of the source candidates of low $S/N$. For ALFALFA data, we estimate the reliability level to be better than $95\%$ for candidate detections with $S/N>S/N_{conf}=6.5$. These are immediately published as cataloged detections \citep[a first catalog of this kind is presented by][]{g07}). Candidate signals with $S/N_{prob}=4.5<S/N<S/N_{conf}=6.5$ will be retained for future pointed follow-up observations, which should deliver corroborated detections at a rate, per unit of telescope time, equal or better than that at which the blind ALFALFA survey detects sources with $S/N>S/N_{conf}=6.5$.

It is not only critical to develop a tool that will allow for the reliable detection of galaxies in the ALFALFA data cubes down to the smallest flux levels possible, it is just as important to understand the properties of such a tool in order to assure the highest quality and completeness for the data catalogs released and to optimally design the confirmation follow-up observations. Since ALFALFA is by far the most sensitive blind HI large-scale survey of the nearby Universe, some of its most interesting and novel scientific results will come from the very lowest mass galaxies and otherwise faint objects it is detecting. The strategy for signal extraction proposed here will capitalize on this strength of ALFALFA by allowing for the reliable detection of such interesting extragalactic systems. 

\acknowledgements
It is a pleasure to acknowledge the many contributions made by Riccardo Giovanelli to this work, including most valuable insights throughout the redaction of this paper. Many thanks also to Martha Haynes, Brian Kent, Sabrina Stierwalt and Rebecca Koopmann, who provided helpful comments during the development and testing phases of the signal extractor software. This work was supported by NSF grants AST-0607007 and AST-9397661, and by grants from the {\it Fonds qu\'{e}becois de la recherche sur la nature et les technologies} and from the Brinson Foundation.

\end{document}